# Adaptable Symbol Table Management by Meta Modeling and Generation of Symbol Table Infrastructures

Katrin Hölldobler [*]     Pedram Mir Seyed Nazari     Bernhard Rumpe

Software Engineering, RWTH Aachen University, Germany

{hoelldobler,nazari,rumpe}@se-rwth.de

**Abstract**

Many textual software languages share common concepts such as defining and referencing elements, hierarchical structures constraining the visibility of names, and allowing for identical names for different element kinds. Symbol tables are useful to handle those reference and visibility concepts. However, developing a symbol table can be a tedious task that leads to an additional effort for the language engineer. This paper presents a symbol table meta model usable to define language-specific symbol tables. Furthermore, we integrate this symbol table meta model with a meta model of a grammar-based language definition. This enables the language engineer to switch between the model structure and the symbol table as needed. Finally, based on a grammar annotation mechanism, our approach is able to generate a symbol table infrastructure that can be used as is or serve as a basis for custom symbol tables.

***Categories and Subject Descriptors***    D.2.11 [*Software Engineering*]: Software Architectures—Domain-specific architectures

## 1. Introduction

Developing a (domain-specific) modeling language or general purpose language involves a multitude of design decisions including the concrete concepts it should be capable of. Textual languages are usually defined by a grammar which results in a tree like structure of the models internal representation, the abstract syntax tree (AST).

[*] K. Hölldobler is supported by the DFG GK/1298 AlgoSyn.

Most textual languages share some common concepts. Typically, a language allows the user to define model elements (resp. program entities [15]) and refer to those from the same model or from different models. For instance, a type of a field in a Java class can refer to another Java class. Furthermore, in many languages some kind of import mechanism provides access to elements of other models. Moreover, some languages provide hierarchical structured elements which enable shadowing names or using identical names for different kinds of elements. For example, fields and methods in Java may have the same name even within the same class.

Handling references and visibilities requires some kind of resolving mechanism that can either be realized by the underlying AST or handled by an additional structure such as a symbol table [1]. A symbol table can be a simple name-information mapping or even a more elaborate data structure that includes the semantic model [3] and can even be used for black box integration of models [12]. However, developing an additional structure can be a tedious task that leads to an additional effort for the language engineer.

Thus, this paper presents an approach to ease the creation of language-specific symbol tables. Therefor, we defined a meta model for symbol tables containing first-level classes for, among others, reference and visibility concepts. By designing this at the $M_3$ meta level, on the subjacent meta level a concrete instance of this symbol table can be created and is fully typed.

Furthermore, whether the AST or the symbol table is better suitable depends on the task to be done, such as generating code, validation, and model integration. In order to enable switching between these different data structures, we integrate the meta models of the symbol table and grammar at the $M_3$ level.

Finally, we adapted the existing annotation mechanism of the MontiCore grammar format [11] to be able to generate a completely systematic symbol table as an instance (i.e., $M_2$ model) of the symbol table $M_3$ model. This can either be used directly or serve as a basis for the creation of a custom symbol table.

In sum, this paper's contribution is (1) a language-independent meta model for symbol tables that allows defining language-specific symbol tables, (2) an integration of this meta model and the grammar meta model, (3) a naming convention for annotations of grammar elements using the example of a MontiCore grammar and (4) based on it the generation of a symbol table that can be used as is or serve as a basis for custom symbol tables.

In the following we first explain the different meta levels involved in this approach in Sect. 2 and examine the grammar meta model in Sect. 3. Thereafter, the symbol table meta model is described in Sect. 4, while in Sect. 5 the integration of both meta models as well as the symbol table generation is explained. Finally, related work is discussed in Sect. 6 and the paper is concluded in Sect. 7.

## 2. Meta Modeling

Our approach of a meta model for symbol tables in conjunction with the generation of a completely systematic symbol table acts on the different meta modeling levels, thus, in this section we give a brief overview of meta modeling levels and clarify which meta level is meant by $M_0$, $M_1$, $M_2$ and $M_3$.

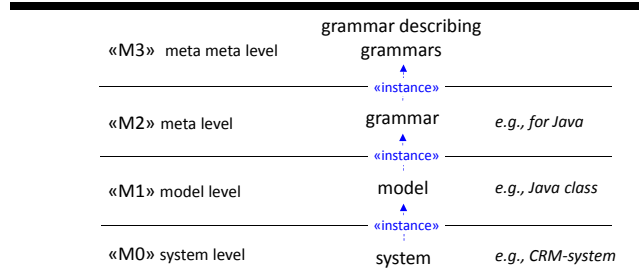

**Figure 1.** Overview of the Different Meta Levels

In meta modeling a distinction is made between different levels/layers of meta modeling referred to as $M_0$, $M_1$, $M_2$, $M_3$, etc., where every level $M_n$ is considered as an instance of the level $M_{n+1}$ [2, 14]. The lowest meta level is $M_0$ which is the real world system (cf. Fig. 1). As our approach targets language design the lowest instance level ($M_1$, model level) considered here is a concrete model, e.g., a Java class according to a language, e.g., Java. In grammar-based approaches, a model is an instance of a language described by a grammar which is the next meta level ($M_2$, meta level)[1]. Finally, the topmost meta level ($M_3$, meta meta level) is a grammar able to describe grammars. In sum, a meta model of a model, is a grammar, and a meta model of a grammar is a grammar of grammars.

## 3. Grammar Meta-Model

Using MontiCore [11] a modeling language is defined by an EBNF-like grammar. A simplified meta model[2] of a Monti-

---

[1] Please note that the $M_2$ model corresponds to the AST.

[2] We deliberately omitted different production kinds and classes representing alternatives etc. to narrow the meta model to parts relevant for the presented approach.

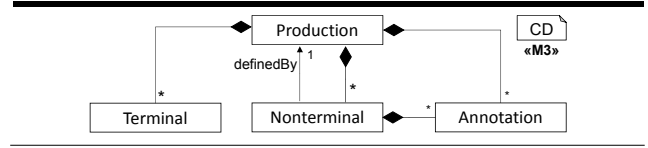

**Figure 2.** Grammar Meta Model (M3 Level)

Core grammar in form of a class diagram is shown in Fig. 2. Thus, a grammar consists of a set of productions. Each of which consists of an arbitrary number of terminals and nonterminals. Similar to EBNF, every nonterminal is defined by one production. Additionally, each grammar element (not shown for `Terminal`) can be annotated with further information, e.g., for documentation purposes.

An excerpt of a simplified meta model for the Java programming language is shown in Fig. 3. This meta model is an instance of the meta model of a grammar shown in Fig. 2. However, for readability reasons we omitted all terminals, chose different names for nonterminals and their defining productions (J-prefix for nonterminals is omitted) and denoted the instance relation in form of stereotypes, e.g., <<Nonterminal>>. The `JClass` production in Fig. 2 represents a Java class. It consists of several `Method` and `Field` nonterminals. `Field` is defined by the production `JField` that represents a Java field or variable declaration, while `Method` is defined by `JMethod` that represents a Java method. A `JField` has a `Type` nonterminal defined by the `Name` production. Furthermore, a `JMethod` consists of several `While` nonterminals defined by `JWhile`. `JWhile` represents a while loop which, among others, consists of fields.

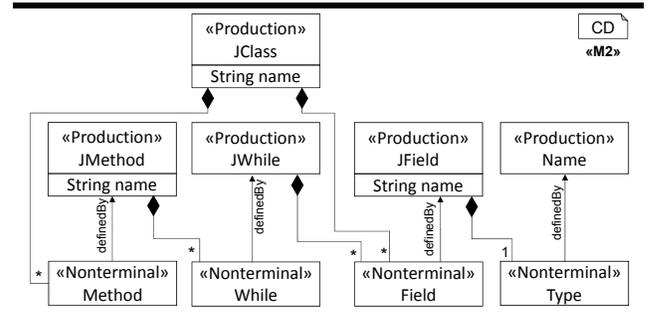

**Figure 3.** Java Meta Model (M2 Level)

## 4. Symbol Table Meta-Model

Many software languages share same or similar concepts, such as: a) The possibility to define and reference model elements. b) References to elements of the same model as well as of another model are allowed. The latter includes the (re-)loading of models. c) The ability to shadow names that are already defined. d) Import statements to enable the use of simple names.

Often, these tasks are conducted by so-called symbol tables (see Sect. 4.1). In the following, we will present the symbol table $M_3$ model of MontiCore. For this, we introduce some core concepts that are common in many lan-guages by taking the example of Java and present the corresponding Java symbol table $M_2$ model.

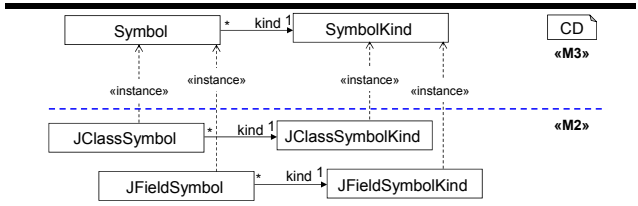

**Figure 4.** M2/M3 Classes for Symbols and Symbol Kinds

Many of the presented concepts and mechanisms are complex and must usually be fully understood by the language engineer in order to apply them. Therefore, the aim of the $M_3$ model is to encapsulate the complexity within the framework and enable the generation of reasonable defaults for a concrete language on the $M_2$ level (see Sect. 5).

### 4.1 Symbols and Symbol Tables

In general, languages have different kinds of model elements, e.g., classes, methods, and fields in Java, each of which has its specific information. Java classes, for example, can be abstract or final and may sub-class other classes. A method can define, among others, a parameter list and a return type. The model elements are represented by a symbol. A *symbol* contains all essential information about a named model element and has a specific *kind* depending on the model element it denotes. Additionally, a symbol can provide information that *is not directly part of the model element*, but useful for the language engineer (resp. generator engineer). For example, a symbol representing a Java class could provide easy access to all non-private fields and methods of all its direct and indirect super types. A *symbol table* (ST) is a data structure that maps names to symbols. It allows to effectively organize and find declarations, types, signatures, implementation details etc. associated with a symbol (resp. model element). A ST consists of a scope-tree (see 4.2) with an associated collection of symbols at each scope.

The $M_3$ model for symbols and symbol kinds is shown in the top part of Fig. 4. The $M_3$ class Symbol has exactly one SymbolKind, whereas a SymbolKind can belong to several Symbols.[3] The corresponding (shortened) $M_2$ model for Java is presented in the bottom part of Fig. 4. The $M_2$ class JClassSymbol is an instance of the $M_3$ class Symbol. Its associated kind JClassSymbolKind is an instance of the $M_3$ class SymbolKind. Analogously, JFieldSymbol and JFieldSymbolKind are instances of Symbol and SymbolKind, respectively. Same applies to methods (not shown).

### 4.2 Scopes

A *scope* holds a collection of *symbol definitions*. In Java, for example, methods and fields are defined in a class scope. Scopes are structured hierarchically, i.e., they can have a direct enclosing scope and several sub-scopes. The resulting structure is a scope-tree (resp. scope-graph) modeled by the `enclosing-subs` association of the $M_3$ interface Scope in Fig. 5.

---

[3] Representing a symbol kind by its own class simplifies the integration of heterogeneous models [5]

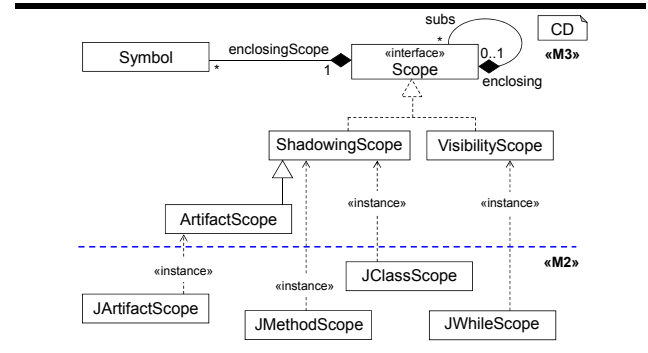

**Figure 5.** M2/M3 Classes for Scopes

*Symbol Visibility* Scopes limit the *visibility* of symbols, i.e., the logical region where the symbol is accessible by its name. For instance, a local variable in Java is only visible within the method scope it is defined in. Outside the method, the local variable is "out-of-scope".

The visibility of a symbol can be *shadowed* by other symbols usually contained in sub-scopes. In Java, a local variable v shadows a same-named field of the class scope. Thus, using the name v *from within the method* refers to the local variable, but *from outside the method* (and in the same class) it refers to the field. However, whether a symbol is shadowed depends on the scope in which the symbol is defined. A Java while-block, for instance, *may not* declare a new variable v, if its enclosing method already does. Consequently, we can distinguish between two types of scopes: *Shadowing scopes* may shadow names that are already defined in their enclosing scope(s) whereas *visibility scopes* may not. In Java, class scopes, method scopes, and artifact scopes (see below) are shadowing scopes. All other scopes are visibility scopes.

Fig. 5 shows the $M_3$ classes for shadowing and visibility scopes and the corresponding Java $M_2$ classes. ShadowingScope and VisibilityScope implement the $M_3$ interface Scope. Consequently, every scope in $M_2$ is either a shadowing scope or a visibility scope. Fig. 6 shows how the enclosing-subs relation stated by the $M_3$ interface Scope is realized for Java on the $M_2$ level. A JClassScope is the enclosing scope of JMethodScopes and JClassScopes (of its inner classes) which are its sub-scopes. The enclosing scope of a JWhileScope (i.e., the scope of a while-block) is a JMethodScope (or another block not shown here). Consequently, a JMethodScope can have JWhileScopes as sub-scopes.

*Artifact Scopes* Models of textual languages usually are stored in an artifact, e.g., a file. Besides the top level element(s), the artifact often contains information about packages (resp. name spaces) and import statements which are important for name qualifying and inter-model references

25

(see Sect. 4.5). In Java, the package and import statements are declared outside the class definition, inside the artifact scope. The *artifact scope* represents the scope of the artifact (resp. compilation unit). It is the top scope of all symbols defined in an artifact and by default a shadowing scope. Thus, `ArtifactScope` sub-classes `ShadowingScope` as shown in Fig. 5. The artifact scope in Java is represented by the $M_2$ class `JArtifactScope` which is an instance of `ArtifactScope`. `JArtifactScope` can only contain class scopes (cf. Fig. 6). Usually, it contains exactly one class scope, but it is also possible to define more than one class in a Java file.

### 4.3 Scope Spanning Symbols

Symbols that span (i.e., define) a scope themselves are called *scope spanning symbols*. A Java class symbol, for instance, spans a scope to enable field and method definitions within that scope. Fig. 7 shows that the `JClassSymbol` is not just an instance of `Symbol` (as shown in Fig. 4) but strictly speaking an instance of `ScopeSpanningSymbol`. It spans a shadowing scope, namely a `JClassScope`. Analogously, `JMethodSymbol` is a scope spanning symbol spanning a `JMethodScope`.

Please note that on the $M_3$ level the relation between `ScopeSpanningSymbol` and `Scope` has the cardinality `0..1`-to-`1`, which means a `Scope` may *optionally* be spanned by a symbol. However, on the $M_2$ level a scope is either always spanned by a symbol or never. For example, a `JClassScope` is always spanned by a `JClassSymbol` since the cardinality is `1`-to-`1`. In contrast, a `JWhileScope` is never spanned by a symbol, as no association to a symbol exists (i.e., `0`-to-`0`).

### 4.4 Symbol References

*Symbol references* refer to symbols that are defined elsewhere, e.g., in other scopes. A symbol definition exists *exactly once* and is stored in a scope. In contrast, *several* symbol references may exist, which are managed in the referencing symbol. In Java, for example, a class `C` *refers* to its super class `S`, since `S` is *defined* elsewhere, e.g., in another file. Fig. 8 shows the corresponding $M_3$ classes. The `definition` association has the cardinality `0..1` instead of `1`, since a `SymbolReference` could refer to a non-existing symbol. Also, it should be possible to load the corresponding symbol definition lazily. The bottom part of Fig. 8 shows an example of a symbol reference for a Java class.

`JClassSymbol` represents the *definition* and `JClassSymbolReference` its *reference*. By referring to `JClassSym-`

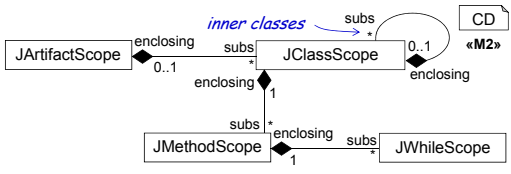

**Figure 6.** Relation of Java Scopes

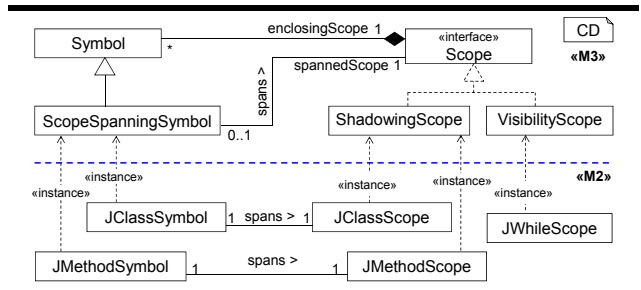

**Figure 7.** M2/M3 Classes for Scope Spanning Symbols

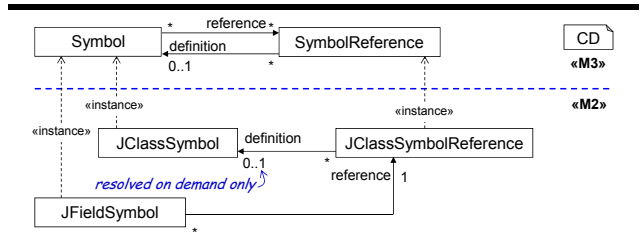

**Figure 8.** M2/M3 Classes for Symbol References

`bol`, `JClass-SymbolReference` can *delegate every request* to the symbol definition. For that, the symbol reference contains all information needed to resolve the corresponding definition, usually, the name and kind of the symbol definition. A field in Java always has a type, hence, `JFieldSymbol` has a `JClassSymbolReference`. It is important to separate symbol definitions and references into different classes since the references can contain additional information that is *specific to the reference*. For example, the field `List<String> l` has a reference to `List` with the type argument `String`. Since other type arguments are possible, such as `List<Integer>`, `List<Boolean>`, etc., it is important to store the information about the type arguments in the symbol reference.

### 4.5 Symbol Resolution Mechanism

Finding the definition of a symbol and the information associated with it is called *symbol resolution* (cf. name resolution [13]). To resolve a symbol usually its name and kind are needed [17]. The symbol kind is needed in the resolving process since many languages allow to use the same name for different elements, e.g., in many object-oriented languages fields and methods may yield the same name.

Resolution algorithms often are very complex and rely on several (language-specific) aspects, such as shadowing, visibility and accessibility rules. Furthermore, these rules can differ depending on the scope level, e.g., Java has different shadowing rules for method and while-blocks. However, many languages share some common resolving mechanisms: a) The starting point is the innermost scope [1]. The resolution continues with the enclosing scope until the symbol definition is found. b) Name occurrences in *shadowing scopes* shadow the ones of same symbol kinds in enclosing scopes. c) Same names may be used for different symbol kinds, e.g., field and method. d) Some import mechanisms are used —usually in the artifact scope— to resolve elements from outside the model.

As described previously, we introduced those concepts on the $M_3$ level, such as `ShadowingScope`, `VisibilityScope`, `ArtifactScope`, `Symbol`, `SymbolKind`, and the corresponding enclosing-subs relations. This enables us to define a resolution algorithm once on the $M_3$ *level* and apply it for every language on the $M_2$ *level*, by using language-specific elements, e.g., `JClassScope`, `JMethodScope`, `JArtifactScope`, `JClassSymbol`, `JMethodSymbol`, `JClassSymbolKind`, etc. In order to match language specific requirements that are not covered by the default behavior, MontiCore provides specific extension points (see Sect. 5.2).

## 5. Integrating and Generating Symbol Tables

The language engineer usually needs both the $M_2$ model of the specific grammar and its corresponding symbol table $M_2$ model. For this reason, we connect the respective $M_3$ models, and hence, enable the composition of the $M_2$ models [7]. Moreover, the language engineer obtains all necessary information about a model element. For example, `JClass` contains syntactical information about a class production. Since it is related to `JClassSymbol`, all further information such as the super class and the members are available in a moderate way. Furthermore, the language engineer need not deal with (re-)loading of referenced models (e.g., the super class). The whole mechanism is encapsulated in the underlying symbol table infrastructure. Thus, on $M_1$ level every processed model provides information about its AST nodes and symbol table elements.

In the following we describe (1) how the two $M_3$ models and the corresponding $M_2$ models are composed, (2) how the composition is conducted syntactically, and (3) how parts of the symbol table infrastructure can be generated by using grammar annotations.

### 5.1 Composing the Grammar and Symbol Table Meta Models

As described in Sect. 4.1, a symbol represents an essential model element. Since those model elements syntactically are defined by grammar productions, we connect the $M_3$ classes `Production` of the grammar and the `Symbol` of the symbol table (see Fig. 9). Note that a symbol always "knows" its kind and its spanned scope (if it is a scope spanning symbol). Hence, it is sufficient to relate a production to the symbol only and obtain the other dependencies transitively. This simplifies the integrated meta models and reduces potential inconsistencies, e.g., if a scope spanning symbol is related to a production, but its spanned scope is not. For the $M_2$ models of the Java example this implies that the productions `JClass`, `JMethod`, and `JField` are related to `JClassSymbol`, `JMethodSymbol`, and `JFieldSymbol`, respectively. We use a `*` cardinality for the relation between `Production` and `Symbol` for two reasons. First, not every production has an associated symbol and vice versa. The `Name` production, for example, is not represented by a symbol. Secondly, a production can define several model elements each represented by a dedicated symbol. For example, a production `JClassMember` could define both, a field and a method that have the corresponding symbols `JFieldSymbol` and `JMethodSymbol`, respectively. Analogously, a symbol `JClassMemberSymbol` can represent the production `JField` as well as `JMethod`.

A `Production` can also be related to a `Scope` without a corresponding symbol. The `JWhile` production, for example, is associated with `JWhileScope`. Again, a `*` cardinality between `Production` and `Scope` is needed, since, for example, a production for an if-else block might be mapped to two scopes.

`Nonterminals` are associated with `SymbolReferences`. For example, the nonterminal `Type` is contained in the `JField` production (see Fig. 3) which itself is associated with the symbol `JFieldSymbol`. Consequently, relating `Type` and `JClassSymbolReference` entails that a `JFieldSymbol` refers to a `JClassSymbol` using the class `JClassSymbolReference`.

### 5.2 Generating the Symbol Table Meta Model

The composition of the two $M_2$ models is affected by the grammar design as well as the symbol table design, which are both determined by the language engineer. The grammar can be designed with just *one production* describing the whole model right up to *many small productions* for every model aspect. Similar, the symbol table can consist of *only one symbol* or *several symbols* for each model element. As a consequence, the composition of the grammar and the symbol table at the $M_2$ level must be conducted manually.

In our experience, there often exists a dedicated production for each *essential* model element. Hence, a `Production` is related to *at most one* `Symbol` and vice versa. The same holds for `Production` and `Scope` as well as `Nonterminal` and `SymbolReference`. In such cases, we can assist the language engineer not only in composing the two $M_2$ models, but also in developing the symbol table using a generative approach.

As described in Sect. 3, the meta model for the grammar is the AST class diagram. MontiCore provides an extended grammar that enables to describe and systematical derive both the concrete syntax and the abstract syntax of a language. A comprehensive description of the MontiCore grammar is given in [11].

We now go one step further and automatically derive besides the concrete and the abstract syntax, the symbol table from the grammar. As stated before, this is only possible to a certain extent, since the language engineer determines the abstraction level of the symbol table. Furthermore, the sym-bol table might contain information that is not directly contained in the grammar (see Sect. 4.1), and hence, it cannot be derived automatically from it. However, in many cases at least the infrastructure of the symbol table can be derived automatically. In the following, we describe this approach and show how the language engineer can make use of it.

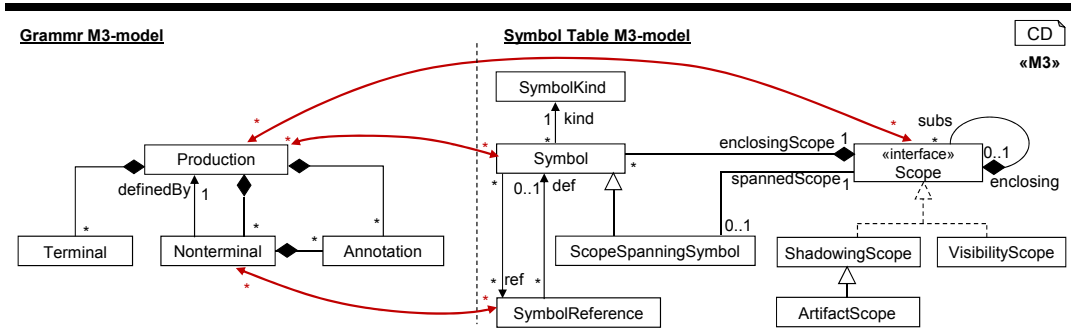

**Figure 9.** Composed M3 Models of Grammar and Symbol Table

We use the existing annotation mechanism of the MontiCore grammar in order to (1) automatically derive the language specific (i.e., $M_2$) symbol table infrastructure from it and (2) simultaneously integrate it with the language-specific grammar $M_2$ model. Annotating a production with

```
JClass@! = "class" Name "{" (JField | JMethod)* "}";
JField@! = type:Name@JClass Name ";";
```

**Listing 1.** Simplified Java Grammar Excerpt with Annotations

`@!` specifies that this production is related to a symbol. The mapping is conducted by a naming convention: a production Prod is mapped to a symbol Prod*Symbol*. In Lst. 1 both productions JClass and JField are annotated with `@!`, hence, they are related to the symbols JClassSymbol and JFieldSymbol, respectively.[4]

By solely marking the productions with an annotation, the two points mentioned above are fulfilled for $M_2$ symbol classes. First, the symbols JClassSymbol and JFieldSymbol can be generated along with their kinds JClassKind and JFieldKind, respectively. Second, the grammar elements and the symbol table elements are related to each other, e.g., JClass and JClassSymbol.

Furthermore, the following aspects can be derived from the grammar *without* being explicitly defined. JClass contains the nonterminal JField which is defined by the same named production that in turn is related to a symbol. Simply put, the JClassSymbol contains JFieldSymbols. Thus, JClassSymbol is a scope spanning symbol. So, its scope JClassScope is generated, too.

A Java field has a type that refers to a class defined elsewhere. Syntactically, the type reference is just a name as stated by the nonterminal type:Name used in the JField production (l. 2, Lst. 1). The nonterminal's annotation `@JClass` specifies that a JClass production is referenced. Again, we choose a naming convention: if a nonterminal is annotated with `@Ann` and Ann is the name of a production, the nonterminal will be related to a symbol reference Ann*SymbolReference*.

From this information we can infer that JFieldSymbol refers to a JClassSymbol as its type.[5] The $M_2$ class Type (see Fig. 3) is related to JClassSymbolReference, hence, a JFieldSymbol has a JClassSymbolReference.

```
JWhile = "while" "(" ... ")" "{" JField* ... "}";
```

**Listing 2.** Simplified Production for a while-Block

Lst. 2 shows a (highly simplified) production of a Java while-block. The JWhile production contains, among others, the nonterminal JField, meaning that Java fields (or rather variables) can be defined in it. Since the JField production is related to a symbol (see Lst. 1, l. 2), we can derive —following the *convention-over-configuration* approach— that JWhile spans a scope. Consequently, a corresponding JWhileScope class is generated that can (only) contain JFieldSymbols. By default, a scope not spanned by a symbol is considered to be a VisibilityScope unless the corresponding production contains the Name (resp. name:Name) nonterminal. For example, if no symbol was created for JClass, it would be considered as a shadowing scope, since it has a name and contains JFields. Finally, JArtifactScope is generated, too, since—as stated in Sect. 4—models of textual languages typically are stored in an artifact (resp. file). The symbols and scopes that may be contained in JArtifactScope are determined from the grammar as follows. Beginning from the start production, find the first productions that are related to a corresponding symbol or scope. Those symbols and scopes may be defined in the artifact scope. In the simple example of Lst. 1, JClass is the start production and is also related to a symbol. Consequently, JArtifactScope may only contain JClassSymbols and their spanned scope. As an artifact scope corre-sponds to the file instead of a model element, there is no class resp. production related to a generated artifact scope.

---

[4] Since a symbol represents a *named* model element (see Sect. 4) a context condition check can ensure that only productions containing the Name nonterminal are tagged with @!.

[5] Please note the difference, as a JClass contains a nonterminal *defined by* JField, the JClassSymbol is a scope spanning symbol. Instead, JField does not contain a JClass nonterminal, but only the Name nonterminal.

Using the above mentioned conventions for annotations in the grammar and the derivation rules, a lot of the symbol table's *language-specific* infrastructure (i.e., $M_2$ model) is generated with the default behavior described in Sect. 4.1. The language engineer has the following options: 1) *Customize and extend* parts of the generated infrastructure using the different integration mechanisms as described in [4] without changing the code directly (see next point). 2) Use it or parts of it as *one-shot generation*, i.e., change the code directly. Consequently, changes in the grammar do not affect the code anymore. 3) Use it *unchanged*, if it fits all the requirements. 4) *Ignore it* and develop the symbol table it manually instead.

## 6. Related Work

Classical symbol tables typically are simple hash tables where a key, the identifier, is mapped to the associated information. Using those symbol tables, some possible implementations of nested blocks are the use of (unique) qualified identifiers or nesting symbol tables per block [1]. Furthermore, if two different kinds of model elements may have the same name, e.g., a field and a method in Java, often one symbol table is created per kind. In our approach, the symbol table is rather conceptually a *table*. The underlying infrastructure is a scope-tree containing a *collection* of symbols (similar to [15]). Each symbol encapsulates the identifier and the associated information. Also, we use explicit *symbol kinds* to distinguish different kinds of model elements. This way, same-named symbols with different kinds may be defined in the same scope.

The purpose of our symbol table approach goes further than in classic compiler construction. It is rather a combination of a simple hash table and a meta model for the *semantic model* as described in [3]: "a semantic model is a representation [...] of the same subject that the DSL describes.". Furthermore, it is "based on what will be done with the information from a DSL script.". Since the purpose of a DSL is determined by the language engineer, the semantic (meta) model cannot be created (resp. generated) automatically. For this reason, we support the language engineer by generating parts of the infrastructure and provide mechanisms for customization.

The meta-DSL *name binding language* (NaBL) [10] allows to specify name bindings (resp. name resolution) and scoping rules declaratively. It provides concepts, such as scoping, definition of imports and names, and referencing rules (cf. [13]). The language workbench Spoofax [6] combines NaBL with the syntax definition formalism (SDF) [16]. Since NaBL models are separated from the syntax definition, they can be easily exchanged and adjusted for different compositions. Unlike our approach, Spoofax's symbol table is a global index with qualified identifiers. Also, we do not provide an own DSL, but follow the convention-over-configuration approach by deriving as much information from the grammar as possible and useful.

Model transformation approaches, e.g., as in [8, 9] conduct transformations between a source and a target $M_3$ models in order to make the corresponding $M_1$ models interchangeable. The transformation is processed in three steps. Firstly, the concepts of the $M_3$ models are mapped to each other. This mapping then enables the transformation of the source $M_2$ model to the target $M_2$ model. Finally, with the $M_3$ mapping and the $M_2$ level transformations the $M_1$ level transformations are derived automatically. Same as in our approach, the mapping is conducted on the $M_3$ level. However, since we need to use information of both $M_3$ models on the $M_2$ and $M_1$ levels, we furthermore *compose* the $M_3$ models. Similarly, our grammar $M_2$ model is the *source* model from which the *target* symbol table $M_2$ model is generated. In contrast to the model transformation approach, we also integrate these two $M_2$ models.

## 7. Conclusion

Textual software languages often share common concepts such as defining and referencing model elements of the same model or other models, shadowing already defined names, and limiting variable visibility using some kind of scoping. Those concepts can be realized by symbol tables which enable easy and efficient access to useful information associated with a model and its elements.

In this paper, we presented a $M_3$ model for symbol tables containing first-level classes for the above mentioned concepts. Based on this meta model the language engineer can develop a language-specific symbol table model at the $M_2$ level. As the symbol table information and the grammar information are related and often required in conjunction, we compose the $M_3$ models, and hence, enable a corresponding composition at the $M_2$ and $M_1$ levels.

Typically, the symbol table is handcrafted, since it highly depends on the purpose of the DSL. Also, composing it with the grammar is a manual task conducted by the language engineer. However, in cases where the grammar matches some criteria—such as containing a dedicated production for each essential model element—it can serve as source model for (1) generating a default symbol table $M_2$ model or parts of it and (2) for directly composing the grammar and symbol table $M_2$ models. For this, we use an existing annotation mechanism for grammar elements and follow the convention-over-configuration approach when generating the symbol table. Different extension mechanisms enable extending and customizing the generated symbol tables.

As a next step, we plan to extend the symbol table $M_3$ model in order to match the requirements of a broader range of software languages. Furthermore, we plan to run experiments to determine whether the suggested defaults and their configuration mechanisms are well understandable and help-ful or can be optimized. Currently, the grammar design widely influences the default generation of symbol tables. Therefore, we will examine whether more complex deriving patterns, e.g., the (transitive) dependency between productions, can improve the symbol table generation.